# Single-cell micro- and nano-photonic technologies

Filippo Pisano[1,$], Marco Pisanello[1,$], Massimo De Vittorio[1,2+], Ferruccio Pisanello[1,+,*]

[1] Istituto Italiano di Tecnologia, Center for Biomolecular Nanotechnologies, Via Barsanti, 73010 Arnesano (Lecce), Italy
[2] Dipartimento di Ingeneria dell'Innovazione, Università del Salento, via per Monteroni, 73100 Lecce, Italy
[$] Equally contributing Authors
[+] Equally contributing Authors
[*] Corresponding author: ferruccio.pisanello@iit.it

**Abstract**
Since the advent of optogenetics, technology development has focused on new methods to optically interact with single nervous cells. This gave rise to the field of photonic neural interfaces, intended as the set of technologies that can modify light radiation in either a linear or non-linear fashion to control and/or monitor cellular functions. These include the use of plasmonic effects, up-conversion, electron transfer and integrated light steering, with some of them already implemented in vivo. This article will review available approaches in this framework, with a particular emphasis on methods operating at the single-unit level or having the potential to reach single-cell resolution.

## 1 Introduction

The possibility to control neural activity with light has been known since 1957 [1] and several types of optical radiation have been exploited to this purpose, including ultra-violet (UV), visible, near-infrared (NIR) or femtosecond-pulsed light [2]. However, direct light modulation of neural activity resulted in limited repeatability among different cellular types, in unavoidable influence on neighboring cells and in ambiguous signal cascades, making it difficult to widely use these techniques in neuroscience research. Over the last 15 years, the use of genetic targeting to express light-gated actuators and fluorescent indicators of neural activity solved most of the limitations of direct light stimulation, driving the use of light radiation to study neural systems.
These approaches are globally known as optogenetic methods [3] and nowadays represent an important complement to single-cell electrophysiology. By providing cell-type specificity, optogenetics allows neuroscientists to operate on molecularly-defined subpopulations of neurons at different spatial scales, from single cells to entire brain regions (or entire brains in small animals, such as zebrafish [4]). Beside a constant development of genetic techniques, this has been made possible by an intense research activity on optical and photonic methods to bring and collect light from the brain [5–7]. However, the need of genetically encoded membrane proteins to control and monitor cellular events imposes major restrictions in the translation of optogenetic methods toward clinical treatments. This intrinsic limitation has raised the need of alternative approaches and boosted the exploration of frontier technologies. We refer to this growing field as photonic neural interfaces, including approaches designed for optogenetics studies as well as systems overtaking the need of genetic targeting. The goal of this review is to describe available technologies in this framework, focusing on methods that have been applied (or have the potential to be applied) to single nerve cells.

In a photonic neural interface (PNI) light radiation is manipulated by a linear or non-linear interaction within a micro- or nano-scale photonic system, which can generate different physical phenomena that enable interaction with nerve cells. The photonic system can: *(i)* act as a light modulator changing propagation properties of light, *(ii)* generate a different light radiation (i.e. at a different wavelength by fluorescence emission or non-linear interactions) or *(iii)* transduce light into a different physical cue, including the generation of electrons or ions to interact with excitable cells. A tentative classification of PNIs, based on the photonic system architecture and on the type of interaction it generates with nerve cells, is displayed in Figure 1.

Substrate technologies (**Figure 1A**) exploit the know-how on planar micro- and nano-fabrication approaches to obtain patterned-light stimulation or to generate localized surface plasmon resonances (LSPRs). LSPRs are non-propagating electromagnetic waves generated by electrons oscillations in nanometric metallic structures. Being at the interface between the metal and the surrounding medium, LSPRs are very sensitive to variations in the environment and are therefore widely used in photonic sensors. In the context of planar devices designed to interface with single neurons in culture, LSPR-based systems were employed in several works to monitor neural activity. As described in detail in paragraph 2, single-cell's electrical signals can indeed be monitored exploiting the high sensitivity of LSPRs to local variations of refractive index [8–10], while the high energy density available at the resonance hotspot allows accessing intracellular signals with plasmonic opto-poration [11] and to enhance Raman signal in cultured neurons through plasmon-assisted Surface-Enhanced Raman Spectroscopy (SERS) [12].

In addition, plasmonic hotspots provide for a bi-directional interface based on light matter interaction at the nanoscale. In the case of injectable photonic systems (**Figure 1B**), reviewed in paragraph 3, plasmonic resonances have been used to trigger neural activity. LSPRs can indeed be obtained in metallic colloidal nanoparticles (NPs), which can be functionalized on the surface of nerve cells and used to generate action potential via local depolarization of the cell membrane. In their semiconductor version, instead, colloidal NPs have been employed to both control and monitor neural activity [13]. In fact, triggering of action potentials can be achieved through up-conversion nanoparticles (UCNPs) that transform near-infrared (NIR) radiation in higher-energy blue or green light that is then used to depolarize or hyperpolarize transgenic neurons expressing channelrhodopsin 2 (ChR2) or halorodopsins (Halo), respectively [14]. To monitor neural activity, instead, NPs can be optimized to link fluorescence intensity variations to extracellular electric field time dynamics, allowing for optical readout of electrical signals [13]. However, an important open problem of injectable PNIs is the fate of nanoparticles in the long term, as their risk of toxicity is not fully understood yet.

Besides these two classes of PNIs based on light-matter interaction at the nanoscale level, a third set of approaches is based on implantable photonic systems (**Figure 1C**), discussed in detail in paragraph 4. Implantable PNIs can be realized by either planar or non-planar fabrication techniques, or they can exploit the small size of the implanted probe to elicit action potentials on single cells exploiting the photoelectric effect [15], and they can potentially interface with single cells [16, 17]. The main drawback of implantable PNIs is the unavoidable tissue damage caused by introducing an external system, such as fiber optics, semiconductor shanks or conducting wires, into the brain. To address this issue, several research works therefore focused on engineering implantable photonic systems to minimize tissue reactions, with reduction of size and rigidity widely recognized as primary targets to minimize glial and astrocytes accumulation around the implant [16, 18]. When mechanical and chemical features of the implant enable for chronic experiments (weeks or months), implantable PNIs provide a robust systems for all-optical control and/or monitor of neural activity [19].

On the base of this classification, this article reviews recent works and discusses future challenges in the field of single-cell photonic neural interfaces, with particular reference to technologies applied to excitable cells.

## 2 Substrate technologies

Micro- and nano-patterned substrates represent a diffused approach to realize planar photonic neural interfaces (PNI) with living neurons at the single-cell level. As schematically represented in Figure 2, photonic elements can be divided in (i) off-chip remotely-controlled optoelectronic microdevices, used to modulate neural activity via optical-relay systems, or (ii) on-chip plasmonic antennas in 2-D or quasi-3D configurations, exploited instead to monitor both intra and extracellular potentials. These architectures' main purpose is to enhance the performance of traditional electrophysiological approaches leveraging the precision and versatility of optical stimulation.

Looking at the micro-scale, micro-electrodes arrays (MEAs) patterned on planar surfaces found a natural complement in spatially-resolved optogenetic stimulation. MEAs are widely used in neuroscientific research; however, even with a high-density of electrodes, large-scale MEAs techniques suffer from poor correlations between electrical recordings and anatomical imaging to reconstruct single cell morphology. This drawback is particularly relevant for applications where cells are densely packed, such as, for example, recordings of Retinal Ganglion Cells (RGCs) to investigate function and treatment of the vertebrate retina [20]. To gain single-cell precision with millisecond precise temporal resolution, MEAs systems have been combined with spatio-temporal optogenetic stimulation (**Figure 2A**). Remote reconfiguration of the illuminating wavefront has been obtained using photonic elements such as liquid-crystal spatial light modulators (SLM) [21] or 2D-arrays of miniaturized, high-power LEDs (µLEDs) [22, 23]. Leveraging the toolset of optogenetic probes and advances in optical techniques for optogenetic stimulation [24], these approaches show promise in characterising functional properties of genetically defined subsets of RGCs [23] as well as in devising novel treatments for retinal degeneration [25–28].

Going down to the nano-scale, nano-patterning of planar surfaces has allowed the first pioneering use of surface plasmon resonances (SPR) and localized SPR (LSPR) to interface with nerve cells (**Figure 2B-D**) [10, 29–32]. The physical principle behind that is based on local changes of surface electrons density on the metal as a response to neural cell's electric field dynamic. This generates refractive index change of the metal nanostructures and small wavelength shifts in the position of the resonance peak. Experimentally, this is sensed by monitoring the intensity of forward scattered light or its reflection as a function of time, at a specific wavelength ($\lambda$). If $\lambda$ is chosen in a region with a large slope in the transmission or reflection spectra, high sensitivity to the cell's electric field dynamic can be obtained. This type of sensitivity applied to excitable cells was first shown in 2008 [29] exploiting SPR (**Figure 2B**) on a 50nm-thick thin film deposited on glass. A similar concept was more recently applied by Howe *et al.* [30] to develop a SPR-based microscopy approach. SPR on a 50nm Au film on glass were used to detect cardiomyocyte contraction at single cell level on a wide field of view (FOV) of 300x300µm through reflectivity changes induced by refractive index variation, exploiting a multimodal microscope design that combines structural (epi-fluorescence, TIRM and SPRFM) and functional (through SPR) imaging channels.

The sensitivity of SPR can be further improved by localizing the electromagnetic waves in the so-called Localized Surface Plasmon Resonances (LSPR), see scheme in **Figure 2C**, which are obtained in nano-structured films or 3D structures. This approach was employed by Zhang *et al.* [10] using pancake-shaped Au nanostructures arrays (140nm diameter and 40nm height, with a pitch of 400nm). These structures were realized by electron beam lithography (EBL) on a glass substrate coated with indium tin oxide (ITO), and used as a substrate for rat hippocampal neural cell growth. Cultured neurons where then chemically stimulated by a 50μM injection of glutamate into the artificial cerebrospinal fluid (aCSF), and both single action potentials and spike bursts were observed in the all optical forward scattering signal modulated by the cell's electric field dynamics.

The peculiar feature of LSPR is that they allow selectively interfacing with neural cells in the close proximity of the plasma membrane, since the energy density decays a few tens of nanometers away from the plasmonic hotspot. However, planar LSPR devices for neural interfaces typically rely on electrochemical interactions mediated by the extracellular medium to record or stimulate neurons. This approach offers advantages in terms of scalability and high fabrication throughput, but has no access to the intracellular compartment. In a recent contribution, Messina and colleagues showed that three-dimensional nanoantennas can provide electrical, chemical and optical access to the inside of the cell at precise locations [31]. This multifunctional operation was achieved by 3D hollow nanoantennas (**Figure 3D**) realized by lithographically impressing a photoresist with secondary electrons from Focused Ion Beam (FIB) milling [33]. Upon NIR illumination, the nanoantenna generates a plasmonic hotspot that has been exploited not only to record enhanced Raman spectra in combination with extracellular recordings from micro electrode arrays (MEAs) [11, 12, 34, 35], but also to open up nanopores in the cell membrane, in a process called plasmonic optoporation [31]. This has allowed tightly controlled intracellular delivery of molecules and single colloidal nano-particles [31, 36, 37] as well as simultaneous intracellular and extracellular recordings across multiple locations in the same neural cell [11].

From the works described above, it is apparent that planar devices offer multiple functionalities when interfaced with cell cultures and brain slices, either through optoelectronic integration or exploiting plasmonic effects. Together with the possibility of performing new types of studies for neuroscience research, this sub-field represents an important test platform for the exploitation of plasmonic effects into *in vivo* photonic neural interfaces. In this perspective, colloidal nanoparticles represent an interesting possibility to obtain highly localized electromagnetic field confinement and they are reviewed in next paragraph.

**3 Injectable Nanoparticles**

As mentioned above, optical stimulation and monitoring of neural activity largely rely on genetically expressed actuators and reporters with peak absorption at visible wavelengths [38]. This highly influential approach has, however, two main drawbacks. First, visible light is highly attenuated in brain tissue and second, any application is subordinated to complex and often expensive procedures of genetic engineering. To address the first drawback, researchers have naturally turned to longer, near-infrared or infrared wavelengths. This effort has been supported by an enlarged palette of red-shifted optogenetic actuators [39] as well as by advanced optical strategies to perform multi-photon optogenetic stimulation [40, 41]. Over the past few years, the use of lanthanide-doped nanoparticles to up-convert IR illumination to visible wavelengths (**Figure 3A**) has emerged as a promising approach for remote-controlled optogenetic stimulation of neural activity, as reported by various works. Various works reported optogenetic stimulation of neurons using up-conversion nano particles (UCNPs) in vitro[42–46]. These results prompted the application of UCNPs to optogenetic activation and inhibition *in vivo* [47, 48]. Very recently, this approach allowed functional activation and inhibition in multiple brain structures with transcranial IR stimulation of injected UCNPs [14]. However, while UCNPs are opening up novel experimental paradigms they still rely on genetically expressed photosensitive proteins.

To tackle this second drawback, researchers are investigating the long-known possibility of controlling neural activity with direct optical stimulation. However, direct optical stimulation of neurons suffers from low spatial specificity and rather poor temporal precision, consistently with a number of models that identified the absorption of laser light in the tissue, and subsequent changes in the membrane capacitance, as a driving factor for the case of infrared direct optical neural stimulation [49]. In recent years, several groups suggested that direct optical stimulation of neural activity could be enhanced if exogenous, extracellular light absorbers were used to generate the required temperature changes.

Working at visible wavelengths, Carvalho-de-Souza *et al.* [50] demonstrated that Au nanoparticles (AuNPs) can be conjugated to high-avidity ligands to enable optical control of neurons resistant to washout and with low AuNP concentration, both in cultured dorsal root ganglion (DRG) neurons and acute mouse hippocampal slices. The AuNPs mediate the photo-excitability by converting 532 nm laser light illumination to a heat-induced variation of the membrane capacitance, resulting in cell depolarization upon illumination (**Figure 3B**) [51, 52]. A few years later, Parameswaran *et al.* used coaxial p-type/intrinsic/n-type silicon nano wires (PIN SiNWs), synthesized via an Au NPs-catalyzed chemical vapor deposition, to elicit neural activity in DRG cultured neurons through 532 nm laser light with energy as low as ~17 µJ [53]. Here, depolarization is mediated by the migration of holes and electrons to the p-type core and the n-type shell, respectively, upon light illumination. This causes a net negative current in the electrolyte through the surface state of the SiNWs.

Meanwhile, a neighboring area of research focused on NIR wavelengths to take advantage of the therapeutic window (600-1200 nm) where water and hemoglobin have low absorption thus enhancing penetration depth while reducing photodamage. Migliori *et al.* circumvented the need of photosensitive genetic compounds by converting thermal energy to initiate action potentials in *Hirudo Verbana* neurons using chemically inert carbon micro-particles [54]. Farah *et al.*, instead, developed a technique called PAINTS (photo-absorber induced neural-thermal stimulation) using holographic stimulation at green and infrared wavelengths to illuminate black micro-particles (6 um in diameter) in contact with the target cells. Acting as effective photo-absorbers, the micro-particles induce microscopic thermal transients that, in turn, activate cultured rat cortical neurons in their surroundings [55]. At the same time, Paviolo and colleagues suggested that infrared neural stimulation (INS) mediated by Au-nanorods potentially allows for higher stimulation rates, superior spatial selectivity and diminished energy deposition [56]. They demonstrated that exposure of Au-nanorods (AuNRs) to laser light at 780 nm induces intracellular calcium transients and increases neuronal cell outgrowth [56, 57]. On a converging path, Eom *et al.* elicited neural activity by effectively converting infrared laser stimulation to a local temperature increase through LSPR excited in AuNR [58]. A similar application in cultured rat primary auditory neurons treated with adsorbing silica coated AuNR has been proposed by Yong *et al.* [59].

While the body of research on alternative, genetic-free approaches for optical stimulation of neural activity is growing, little has been reported on monitoring neural activity without genetically-encoded activity indicators. An interesting approach employs semi-conductor nano-crystals or quantum dot (QDs) to visualize cellular membrane potential (**Figure 3C**). Nag *et al.* proposed a bio-conjugate composed of QDs, a peptide for membrane inserting, and fullerene ($C_{60}$) [60]. In this design, the QDs act as donor of electron to the $C_{60}$, lying within the membrane bilayer, with the membrane potential mediating the electron transfer (ET) process and, consequently, the quenching of the QDs photoluminescence (PL): depolarization of the cell enhances the ET and the PL quenching, allowing to track the cellular dynamics with greater ΔF/F when compared with conventional voltage sensitive dyes (20 to 40 folds). The efficacy of such nano-bioconjugate was demonstrated *in vivo* in the mouse parietal cortex undergoing electrical stimulation while acquiring wide-field epi-fluorescence images (at a rate of 200Hz). The optical properties of QDs, such as high two-photon absorption cross-section and resistance to photobleaching, might be beneficial in enhancing patch clamp recording by helping in visualizing the pipette in deep brain layers. As shown by Andrásfalvy *et al.* [61] and Jayant *et al.* [62], the pipette can be made permanently fluorescent and easier to visualize by repeatedly dipping its tip in a CdSe/ZnS QDs-hexane solution.

## 4 Implantable technologies

The combination of optogenetic stimulation and large-scale neural recordings has represented a promising path in brain investigation, with the main goal of interfacing simultaneously with a high number of units at the single cell level *in vivo*. To serve this research drive, the toolset of innovative devices to interface with deep-brain structures has been rapidly populated with implantable probes capable of eliciting and recording neural activity (**Figure 4**). While most of these devices have been designed to address multi cellular compartments, recent advances in micro and nano-patterning of photonic elements have shown potential towards the realization of integrated implantable devices working on multiple single units.

UCNPs technology, highlighted in the previous section as an injectable tool, lies at the core of a micro-scale untethered implantable device for optogenetic stimulation. Wang, Lin, Chen, *et al.* packaged UCNPs in a glass micro-electrode and, in combination with a robotic laser projection system, achieved behavioural conditioning in freely-moving animals by modulating neural activity in various brain regions [63]. An alternative upconversion approach, proposed by Ding *et al.* [64], relies instead on the realization of self-powered microscale devices through the combination of GaAs photovoltaic diodes (PDs) and AlGaInP-based light emitting diodes (LEDs); a thin (9μm) implantable probe for stimulating nerve cells in the mouse primary somatosensory cortex via NIR-to-visible conversion (~810nm to ~630nm) was obtained. Even though the single photodetectors/light source couple is quite large (220μm × 220μm) if compared to neurons size, preliminary exploration to scale down the device to an area of 10μm × 10μm was made [64].

By virtue of their small size, relatively low power consumption, and easiness of integration within conventional planar micro- and nano- fabrication techniques, micro-LEDs (µLEDs) are one of the most used technology for the development of integrated implantable probes for optogenetic *in vivo* neural stimulation. Multipoint light delivery devices have been developed on Sapphire [65, 66] and Si [67, 68] shanks, in combination with planar electrode for extracellular recording. The application of µLED probes at single-cell resolution, however, is hindered by the incoherent lambertian emission of LED sources that allows little control on the beam shape. In addition, LEDs act as a heat source that might have an effect on neighbouring cells [69]. This drawback might be alleviated by probes using micro and nanophotonic circuitry to deliver excitation light from multiple locations, often adjacent to recording electrodes, while placing light sources outside the target tissue. This configuration has been proposed by Schwaerzle *et al.* that incorporated Pt-recording electrodes with µ-waveguides excited with on-chip laser diodes in a Si-based device [70]. The design allows delivering laser light ($\lambda$=650nm) in the vicinity of the electrodes while reducing undesired tissue heating. The reduced cross-section of the waveguides (<20 µm on the larger dimension) is well suited for single-cell excitation. Li and colleagues, instead, obtained straight, off-shank light beams with little angular divergence using nano-patterned gratings to outcouple light from multiple waveguides that are monolithically integrated with a Micro Electrode Array (MEA) built on silicon substrate. Using these probes, they successfully stimulated and recorded neural activity in the mouse secondary motor cortex from single units [71, 72]. Mohanty *et al.* demonstrated *in vivo* control of neural activity with a reconfigurable nanophotonic platform that dynamically modulates multiple light beams with sub-ms temporal control. This was achieved by incorporating multiple interferometric switches to selectively channel a single input into 8 separated waveguides that terminate with a nanograting outcoupler [73]. An alternative nanophotonic probe for spatial-temporal neural control of neural activity has been proposed and demonstrated by Segev *et al.* [74]. In this latter case, spatial selectivity is obtained through a nanophotonic circuit that routes individual spectral components of multispectral light input to specific emitting locations on the shank. The probes leverages arrayed waveguide gratings designed for visible wavelengths and, as in [71, 72], relies on diffractive gratings to outcouple stimulation light off-shank. Monolithic intracortical implants for light delivery and simultaneous electrical recording of activity made by ZnO were developed also by Lee *et al.* [75], as arrays of needle shaped micro optrodes with tip size of few tens of micrometers: the ZnO crystal work both as waveguide for visible light and electrical interface, and each element of the array is independently addressed with light through a laser scanning system once implanted in cortex. Beside solid-state photonics on planar substrates, optical control of guided modes on tapered optical fibers [76–78] allows for engineering the size and the shape of the stimulation volume [79, 80] approaching single cell dimensions, with light that can emerge from multiple and selectable surfaces as small as 100µm$^2$ [81–83].

Together with implantable devices for optogenetic control of neural activity, also the field of functional fluorescence detection from single or few cells *in vivo* has recorded an increasing interest, mostly driven by the limited performance of flat-cleaved optical fibers [84]. In particular two works have been recently presented exploiting both solid-state devices and waveguide approaches. Lu *et al.* [17] developed a wireless photometer integrating a microscale inorganic LED (µILED; 270µm × 220µm × 50µm) and a GaAs micro photodiode (100µm × 100µm × 5µm) on a needle-shaped flexible substrate. Recently, Pisano *et al.* [16] proposed a single waveguide device potentially able to address both wide volume and single cell fiber photometry by exploiting the modal demultiplexing properties of micro-structured tapered optical fibers .

Nevertheless, stiff implantable device like silicon-based shanks or glass waveguides, trigger foreign body reactions resulting in acute and chronic tissue damage [85, 86]. These effects can be mitigated by using flexible and stretchable multifunctional devices, such as the entirely polymeric multifunctional fibers for simultaneous optogenetics stimulation, electrophysiology recording, and drug delivery *in vivo* developed by the Anikeeva lab [18, 87]. These integrated multimodal probes, with a total diameter varying in the range 180μm-220μm, were obtained by heating and stretching a macroscopic size preform and are composed by a polycarbonate-cyclic olefin copolymer waveguide, several graphite-conductive polyethylene electrodes for extracellular recording, and microfluidic channels. Wirelessly interfaced needle-shaped flexible multifunctional devices integrating μILED, miniaturized photodetectors, and fluidic channels have also been realized by the Rogers lab [17, 88, 89].

Recently the attention of the scientific community was also captured by the possibility to functionally image the neural activity in freely behaving animals [90]; nevertheless, imaging neural population in deep brain layers beyond the reach of conventional two-photon microscopy requires the use of high-resolution endoscopic devices with reduced cross-section. To this end, graded index (GRIN) lenses [91] have been widely employed to image neurons in deep structures of the mouse brain with single, multi-photon or patterned multi-photon illumination [92–96]. To further reduce the implant cross sizes, researchers are pursuing wave front modulation techniques to control modal propagation in single multi-mode optical fibers in order to produce tight focal spots at the fiber distal end [97–101]. This ground-breaking approach has recently enabled 3D, sub-cellular resolution imaging of dendrites and dendritic spines in the mice dorsal striatum as well as spatially resolved $Ca^{2+}$ imaging in rat organotypic hippocampal slices with a 50 μm-core optical fiber [102]. On the other hand, innovative micro-fabrication approaches have been recently employed to achieve cellular resolution imaging using a mm-long probe with a high numerical aperture lens monolithically integrated on a polymeric waveguide [103].

All of these devices, however, require genetic modification of neurons to make them express opsins or fluorescent activity reporters. Different approaches, as summarized in **Figure 4B**, have been proposed by the scientific community to obtain implantable interfaces without the need of genetically encoded proteins. Multiphoton photoelectric effect has been exploited to realized NIR-based, opsins-free and free-floating stimulation implants by Stocking *et al.* [15]. Their small (8μm diameter) glass insulated carbon fiber probes exploit multiphoton photoelectric effect to achieve neural stimulation: electrons moved from the valence to conduction band within the surface of the carbon fiber generate a local potential change capable to trigger the opening of voltage-gated ion channels, forcing cells to depolarization. The system was validated *in vivo* in mouse visual cortex, and even though multiple cells were elicited during the experiments spatial selectivity could be improved by accurate delivery of both light and carbon probes. On the other hand, the refractive index dependent wavelength shift of SPR of an Au film deposited on optical fibers can be exploited to record neural activity *in vivo* without the needs of $Ca^{2+}$ or voltage fluorescent reporters [104, 105].

**5 Conclusions**

The field of photonic neural interfaces for single cells is representing a great complement of existing tools to study neural circuits, providing spatial resolution, cell type specificity and time-resolved control of cellular functions. The development of technologies to optically interface with single or few units was initially drawn by the advent of optogenetics, that fostered applications at visible wavelengths, matching the absorption spectra of light-gated ion channels and pumps. This is the case of the photonic implants for optogenetics summarized in Figure 4A, based on planar waveguides or optical fibers and employed to deliver or collect light from multiple points of the brain. In parallel, the scientific community is also seeking for alternative methods that could overcome the genetic targeting requested by optogenetic methods. Although they are still at their embryonic stage, approaches based on surface plasmon resonances (SPRs) and on localized surface plasmon resonances (LSPRs) are showing their potential not only to control and monitor neural activity, but also to access chemical information in the cellular environments and/or compartments by surface enhanced Raman spectroscopy. SPRs and LSPRs have found application in implantable devices to monitor neural activity (Figure 4B), and in planar systems to record action potentials and to optoporate the plasma membrane (Figure 2B-D). If implantable and planar neural interfaces represent commonly adopted methods, an alternative path is the use of injectable nanoparticles (Figure 3). These approaches exploit the small size of the nanoparticles to enhance plasmonic effects or to engineer their fluorescence properties, representing a new frontier for next generation of optical methods to access brain functions.

**Acknowledgments**
FeP and FiP acknowledge funding from the European Research Council under the European Union's Horizon 2020 research and innovation program (#677683). MP and MDV acknowledge funding from the European Research Council under the European Union's Horizon 2020 research and innovation program (#692643). FeP and MDV acknowledge that project leading to this application has received funding from the European Union's Horizon 2020 research and innovation programme under grant agreement No 828972. MDV is funded by the US National Institutes of Health (U01NS094190). MP, FeP and MDV are funded by the US National Institutes of Health (1UF1NS108177-01).

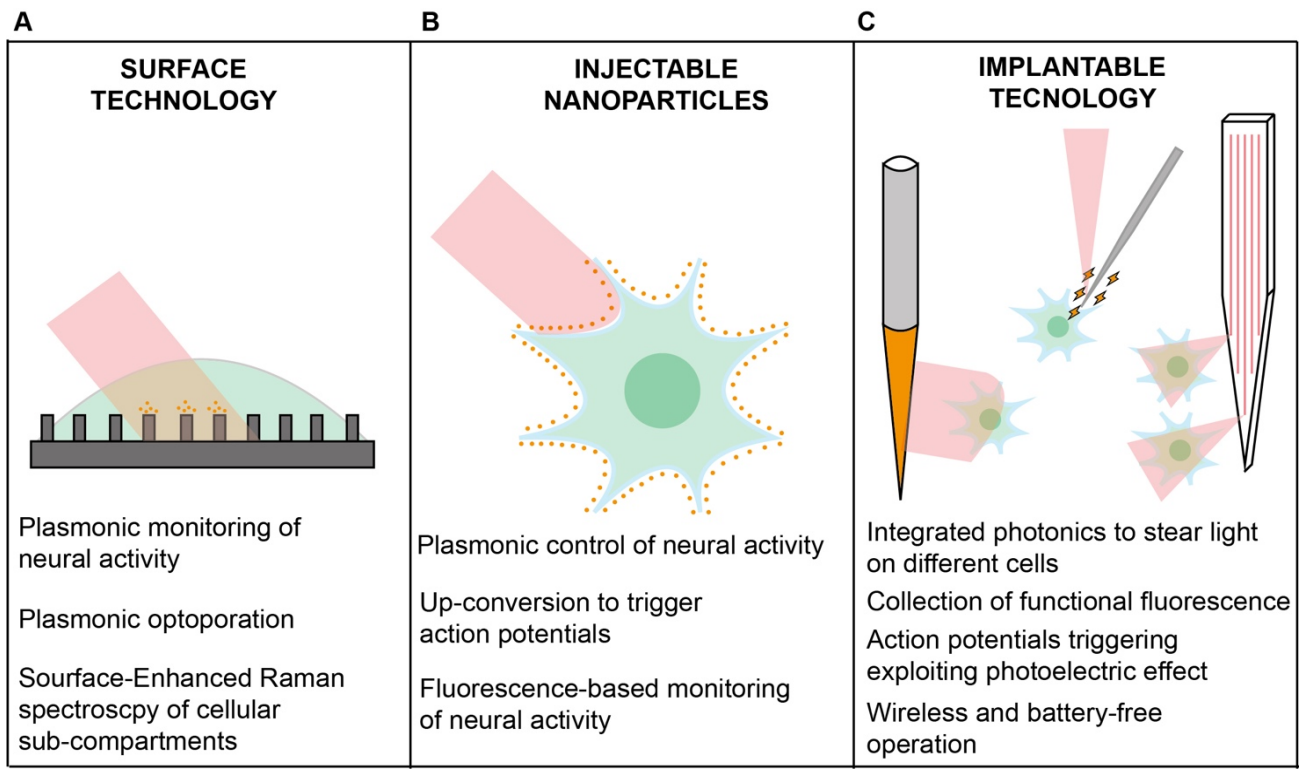

**Figure 1:** A summary of the different approaches developed by the scientific community to realize photonic neural interfaces

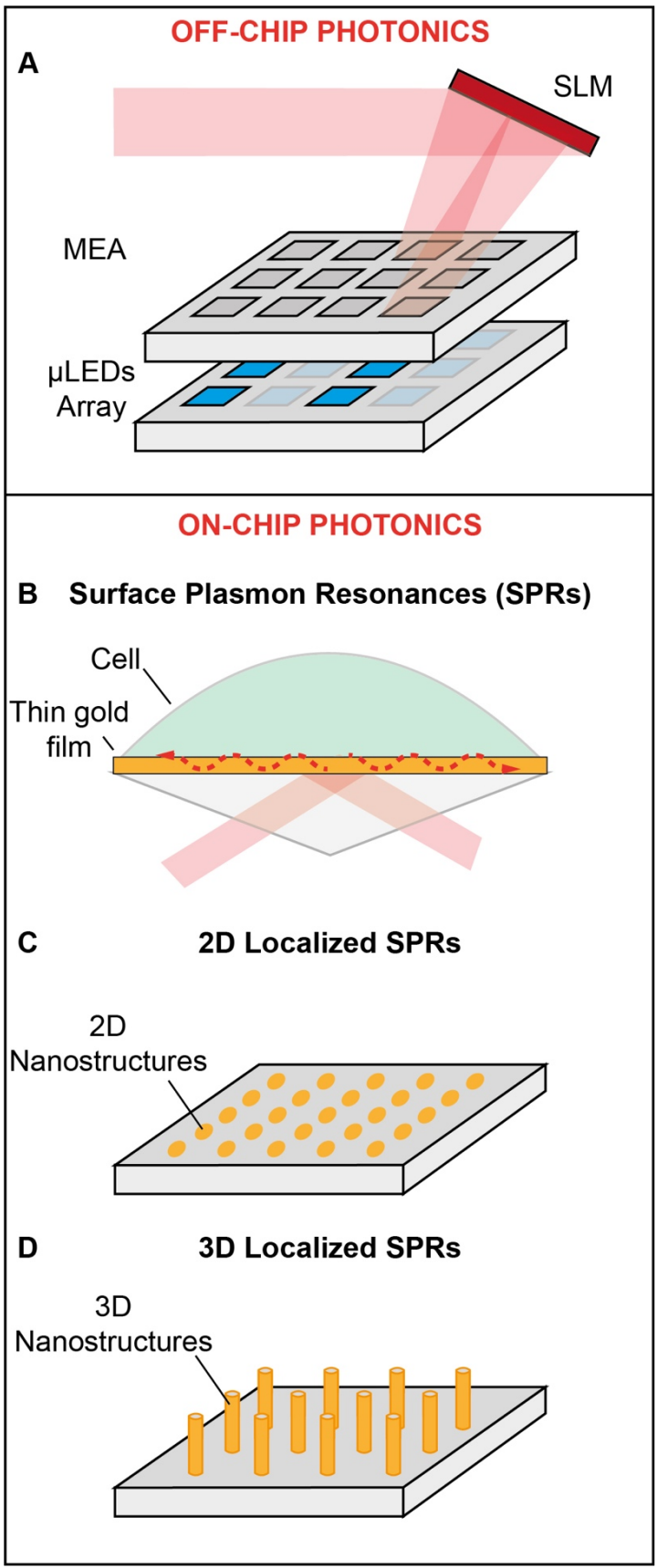

**Figure 2**: Examples of off-chip and on-chip photonics used for planar neural interfaces.

**Injectable Nanoparticles for...**

A  .... Optogenetic Control Neural Activity by Up-Conversion effects

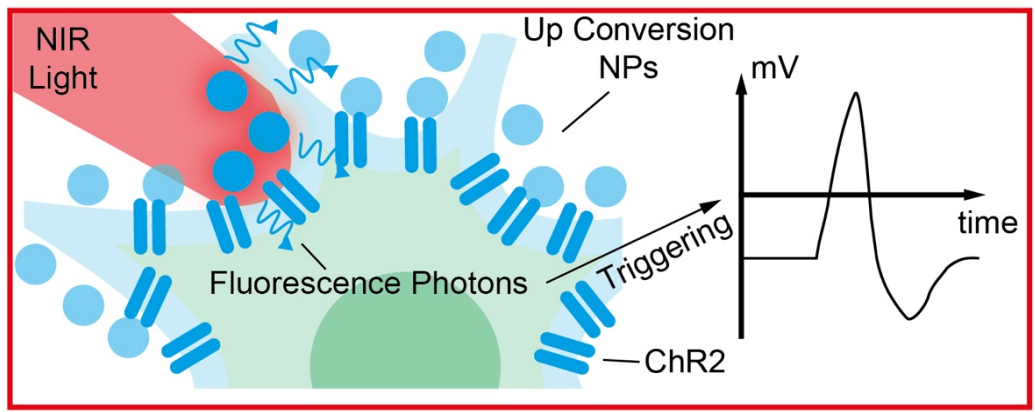

B  ... Triggering Action Potentials by Plasmonic Effects

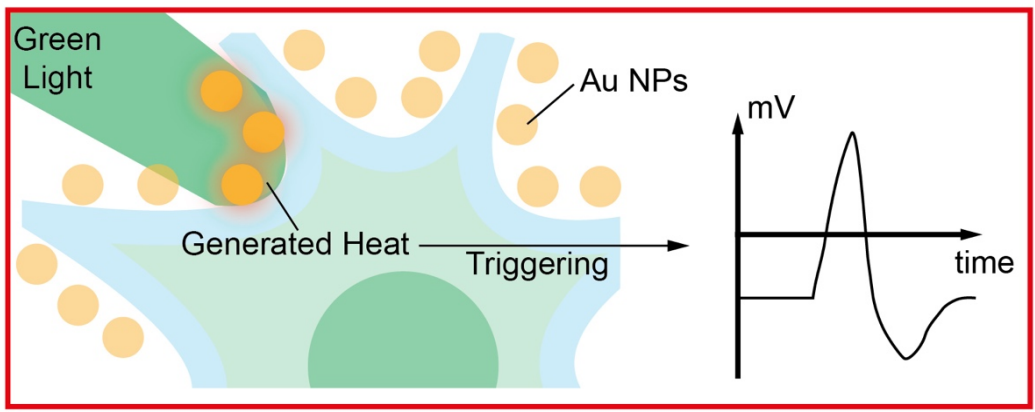

C  ... Monitoring Neural Activity by quantum-confinement effects on fluorescence

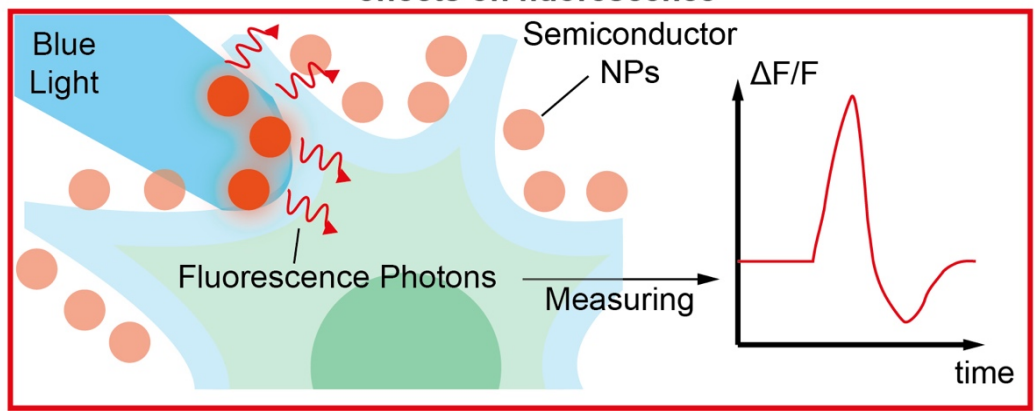

**Figure 3**: Schematic representation of the different approaches used to realize injectable single-cell photonic neural interfaces. (A) Up-conversion nanoparticles, (B) plasmonic nanoparticles and (C) colloidal quantum dots.

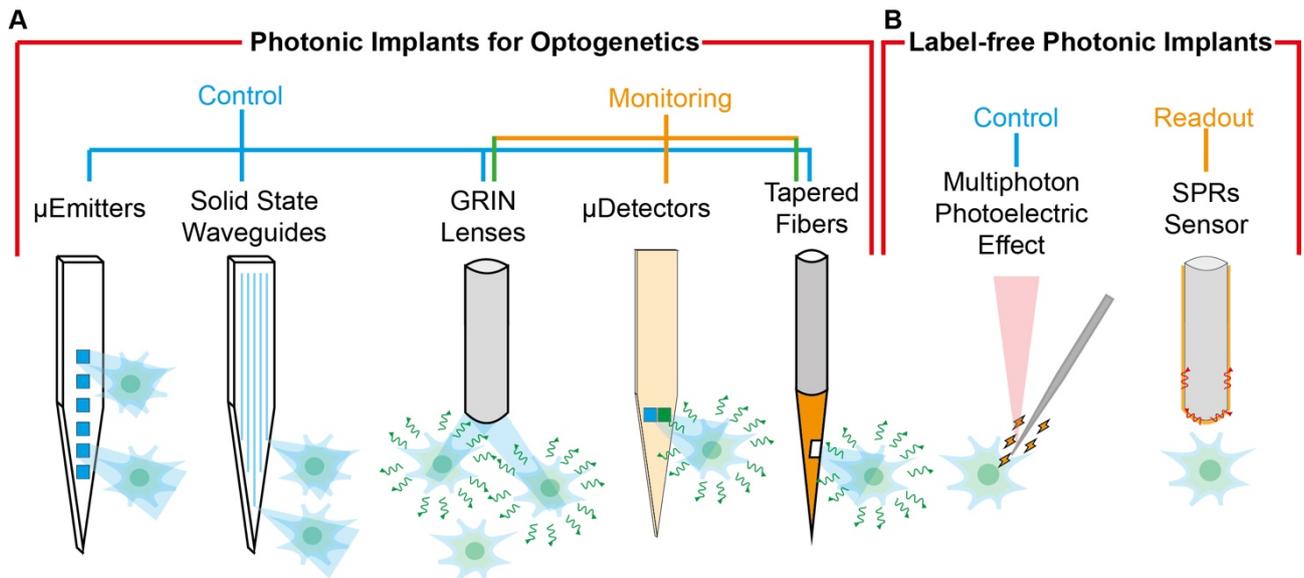

**Figure 4:** Implantable Photonic neural interfaces. (A) Devices for optogenetic control and/or monitoring of neural activity (blue and orange lines, respectively). (B) Label-free approaches.